\documentclass[]{tMPH2e}

\usepackage[utf8]{inputenc}
\usepackage[english]{babel}

\usepackage{graphicx} 
\usepackage{color}
\usepackage{acronym}

\renewcommand{\vec}[1]{\mathrm{\mathbf{#1} } }
\newcommand{\upd}{\mathrm{d}}
\newcommand{\kB}{k_\mathrm{B}}

\begin{document}

% definitions of acronyms
\acrodef{SHM}{shell-bead model}
\acrodef{PC}{plastic crystal}
\acrodef{F}{fluid}
\acrodef{fcc}{face-centered cubic}
\acrodef{BD}{Brownian dynamics}

\acrodef{RDF}{radial distribution function}
\acrodef{PCF}{pair correlation function}

\acrodef{PBC}{periodic boundary conditions}
\acrodef{ACF}{autocorrelation function}
\acrodef{SACF}{stress autocorrelation function}
\acrodef{VACF}{velocity autocorrelation function}
\acrodef{DACF}{directional autocorrelation function}

\acrodef{COM}{center of mass}
\acrodefplural{COM}[COMs]{centers of masses}

\acrodef{DB}{dumbbell}
\acrodef{HS}{hard sphere}

\bibliographystyle{tMPH} 

\title{Equilibrium structure and fluctuations of suspensions of colloidal dumbbells} 

\author{%
	{Nils Heptner and Joachim Dzubiella$^*$}\\\vspace{6pt}%\\
	{Institut f{\"u}r Physik, Humboldt-Universit{\"a}t zu Berlin, Newtonstr.~15, D-12489 Berlin, Germany}\\\vspace{6pt}%
	{Soft Matter and Functional Materials, Helmholtz-Zentrum Berlin, Hahn-Meitner Platz 1, D-14109 Berlin, Germany}\\\vspace{6pt}%
}
\thanks{$^*$To whom correspondence should be addressed. E-mail: joachim.dzubiella@helmholtz-berlin.de}

\maketitle

\begin{abstract}
We investigate the structure and equilibrium linear-response dynamics of suspensions of hard colloidal dumbbells using
Brownian Dynamics computer simulations. The focus lies on the dense fluid and plastic crystal states of the colloids with 
investigated aspect (elongation-to-diameter) ratios varying from the hard sphere limit up to 0.39, which is roughly the 
stability limit of the plastic crystal phase. We find expected structural 
changes with larger elongation with respect to the hard sphere reference case and very localized orientational correlations, 
typically just involving next-neighbor couplings. These relatively weak correlations are also reflected in only minor effects
on the translational and rotational diffusion coefficients for most of the investigated elongations. 
However, the linear response shear viscosity exhibits a dramatic increase at high packing fractions ($\phi\gtrsim 0.5$) 
beyond a critical anisotropy factor of about $L^* \simeq 0.15$ which is surprising in view of the relatively weak changes found before on the 
level of colloidal self-dynamics. We suspect that even for the small investigated anisotropies, newly occurring, collective 
rotational-translational couplings must be made responsible for the slow time scales appearing in the plastic crystal. 
\end{abstract}


\section{Introduction}

Suspensions of hard spherical colloids have been extensively studied theoretically as well as experimentally as a model system for simple liquids in and out-of equilibrium~\cite{HansenMcDonald, Pusey:Nature, LesHouches, Loewen2001, Ackerson1990a, Haw1998}. 
However, even weak anisotropies of the particle shape can have dramatic effects on the structure, dynamics, and phase behavior of colloidal suspensions while  
the details of their molecular origins are not well understood.
Particularly interesting is the question at which elongation what dynamic and structural properties start to deviate from the limiting hard sphere reference case.
To address such a fundamental question, well-defined experimental and theoretical model systems are in need.
One of the simplest realizations of mildly anisotropic colloids are dumbbell-shaped particles (that is,~a dimer of two fused colloidal spheres) which can be synthesized nowadays by different routes in well-controlled and monodisperse ways~\cite{Johnson,Mock2007,Mock2007a, Kramb2010, Kramb2011, Kramb2011a, Forster2011, wagner, Hoffmann2008, Nagao2010, Peng2012, Chu2012}.
The phase diagram of hard-core dumbbells is theoretically known, where, at high packing fractions, the colloids can form the so-called plastic crystal phase~\cite{Ruth, Vega1992, Vega1992a, Vega1997, Marechal2008, Ni2011, Paras1992}. Here, the translational degrees of freedom are essentially frozen at high packings as in the hard sphere  system but the particles are free to rotate. 
Experimental realizations of dumbbell suspensions have been shown to fit well into the respective regions of the theoretical phase diagram~\cite{Chu2012, Chu2014}.
Thanks to these recent advances, viable, purely repulsive monodisperse model systems are experimentally available which crystallize in equilibrium conditions without the help of external fields or long-range interactions. 
Colloidal dumbbells are thus among the most promising experimental model systems for a fundamental understanding of structural and dynamic effects of increasing the elongation with respect to the hard sphere reference case, in particular in the (plastic) crystal phase where new phenomena are expected~\cite{Ruth}.

In this work, we systematically study the structure and fluctuations of  suspensions of steeply repulsive dumbbell particles in equilibrium by means of Brownian Dynamics (BD) computer simulations.
We focus on the dense regime of the \ac{F} to \ac{PC} transition region of the phase diagram which is well accessible for experimental dumbbell suspensions and promises interesting insight into the influence of elongation on steric correlations and structural properties in both fluid and crystal.
Apart from simple spatial correlation functions, we also systematically explore the dynamic properties in the linear response regime at high volume fractions measured by equilibrium fluctuations.
All results for the elongated dumbbells are discussed with respect to the well-explored hard sphere reference system. 

\section{Theoretical methods}
	\subsection{Dumbbell model and Brownian Dynamics}
\begin{figure}[ht]
		\centering
		\includegraphics[]{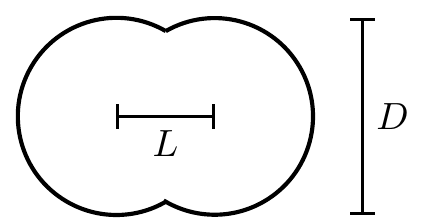}
		\caption{Sketch of the dumbbell geometry with dimensionless aspect ratio $L^\ast = L/D$.}
		\label{fig:model_sketch}
	\end{figure}

	In our model a dumbbell particle consists of two spherical beads of diameter $D$ rigidly constrained at a center-to-center distance $L$, see Fig.~\ref{fig:model_sketch}. With that we define a dimensionless elongation, or, aspect ratio, as $L^\ast = L/D$. For $L^\ast = 0$, we recover the reference case of one spherical colloid.
	
	The site-site (bead-bead) pair interaction is modeled by a Yukawa (or 'screened Coulomb') potential which interpolates between the long-ranged electrostatic Coulomb and the purely excluded volume interaction.
	The potential between two beads at distance $r$ is given by
	\begin{align}
		V\left(r\right) &= \epsilon \frac{\sigma}{r} \exp\left\lbrace -\kappa\left(r- \sigma \right)\right\rbrace.
		\label{eq:potential}
	\end{align}
	The parameter $\kappa$ is a screening constant, which may be used to tune the softness of the pair interaction.
	In the limit of $\kappa \rightarrow \infty$ one obtains the purely hard-core interaction. We choose $\kappa=20/\sigma$ as a compromise between 
	a steep hard-core like repulsion at the same time being smooth enough to not cause numerical problems in the integration. 
	The parameters $\sigma$ and $\epsilon=k_BT$ define the length and energy scale of the interaction in our model, respectively.
	Since we compare our results to the hard-dumbbell and hard-sphere reference system, we define the effective bead diameter $D$ via the Barker-Henderson relation~\cite{Barker1967}
	\begin{align}
		D &= \int\limits_0^\infty \left(1 - \exp\left\{-\frac{V(r)}{k_BT}\right\}\right) \upd r.
		\label{eq:diameter}
	\end{align}		
	The effective volume fraction $\phi = \rho(\pi/6) \alpha(L^\ast) D^3$ at dumbbell number density $\rho$ is defined accordingly. 
	The function $\alpha(L^*) = 1 + 3 L^*/2 - {L^*}^3/2$ is a geometric factor accounting for the bead overlap volume~\cite{Vega1992}.
	
	The total instantaneous force $\vec{F}_{ij}(t)$ exerted by particle $j$ on particle $i$ is the sum of the forces $\vec{F}_{(i,k)(j,l)}(t)$ acting between the $k$-th site of the $j$-th particle and the $l$-th site of the $j$-th particle, via
	\begin{align}
		\vec{F}_{ij}(t) &= \sum\limits_{k,l = 1}^2 \vec{F}_{(i,k)(j,l)}(t).
	\end{align}
	The \ac{COM} vector of the $i$-th particle is decomposed with respect to the axial symmetry as 
	\begin{align}
		\vec{R}_{i}(t) &= \vec{R}_{i,\perp}(t) + \vec{R}_{i,\parallel}(t), \\
		\vec{R}_{i,\parallel}(t) &= \left[\vec{\Omega}_i(t)\cdot\vec{R}_i(t)\right] \vec{\Omega}_i(t),
	\end{align}
	where $\vec{\Omega}_i(t)$ is the unit vector pointing in the direction of the connection between the centers of the beads belonging to the $i$-th particle.
	The total force $\vec{F}_i(t)$ is decomposed analogously:
	The parallel and perpendicular parts are given by
	\begin{align}
		\vec{F}_i(t) &= \vec{F}_{i,\perp}(t) + \vec{F}_{i,\parallel}(t), \\
		\vec{F}_{i,\parallel}(t) &= \left[\vec{\Omega}_i(t) \cdot \vec{F}_i(t)\right]\vec{\Omega}_i(t).
	\end{align}
	
	Given this framework, we implement the \ac{BD} algorithm for the above defined coordinates as follows:
	The particle positions are updated using an explicit forward Euler method with the finite time step $\Delta t \ll \tau$.	
	Starting from the configuration $\{\vec{R}_i^n, \vec{\Omega}_i^n\}$ at time $t = n \Delta t$, we find new coordinates $\left\lbrace\vec{R}^{n+1}_i, \vec{\Omega}^{n+1}_i\right\rbrace$ at $t + \Delta t$ following the scheme
	\begin{align}
		\vec{R}_{i,\parallel}^{n+1} &= \vec{R}_{i,\parallel}^{n} + \Delta t \frac{D_{\parallel} }{\kB T} \vec{F}^n_{i,\parallel} + \delta r_{i,\parallel} \vec{\Omega}^n_{i},\\
		\vec{R}_{i,\perp}^{n+1} &= \vec{R}_{i,\perp}^{n} + \Delta t \frac{D_{\perp} }{\kB T} \vec{F}^n_{i,\perp} + \delta r_{i,1}\vec{e}^n_{i,1} + \delta r_{i,2} \vec{e}^n_{i,2},
	\end{align}
	where $\vec{e}_{i,1}$ and $\vec{e}_{i,2}$ are unit vectors oriented perpendicular to the director $\vec{\Omega}_i$.
	The random variates $\delta r_{i, \alpha},\ \alpha \in \left\{\parallel, 1, 2\right\}$ have zero mean and the variances $\big\langle \delta r^2_{i,\parallel}\big\rangle = 2D_\parallel \Delta t$ and $\big\langle\delta r^2_{i,\lbrace 1,2\rbrace}\big\rangle = 2 D_\perp \Delta t$, respectively. 
 
	The torque acting on symmetric dumbbells is given by $\vec{T}_i(t) = \frac{\sigma}{2} L^\ast \vec{\Omega}_i(t) \times \left(\vec{F}_{(i,2),\perp}(t) - \vec{F}_{(i,1), \perp}(t)\right)$, where $\vec{F}_{(i, k), \perp}(t)$ is the total instantaneous force on the $k$-th bead of the $i$-th particle.
	The internal director of the $i$-th particle is updated according to
	\begin{align}
		\vec{\Omega}_i^{n+1} &= \vec{\Omega}_i^{n} + \Delta t \frac{D_r}{\kB T} \vec{T}_i^n \times \vec{\Omega}^n_i + \delta x_1 \vec{e}_{i,1}^n + \delta x_2 \vec{e}_{i,2}^n,
	\end{align}
	where $\delta x_1$ and $\delta x_2$ are zero mean Gaussian distributed random variates with variance $\left<\delta x_j^2\right> = 2 D_r \Delta t$.
	The single particle mobility is given by parallel $D_\parallel$, perpendicular $D_\perp$ and rotational $D_r$ diffusion coefficients.
	We neglect hydrodynamic interactions between the colloids which is justified for the high packing fractions where steric correlations dominate the equilibrium structure and linear response behavior~\cite{Heyes1994,Phung1996}. 

	The infinite-dilution (single-particle) self-diffusion coefficients of a dumbbell particle were calculated previously using the \ac{SHM} \cite{Carrasco1999,delaTorre2002,delaTorre2007,Diaz1989,Tirado1984}.
	In this method, the particle surface is represented by a number of mini-beads which act as sources of hydrodynamic friction, where no-slip boundary conditions are assumed.
	The mobility tensors are calculated on the Oseen-level using an increasing number of mini-beads.
	Using these tensors, the values are extrapolated to zero mini-bead size, e.g., infinite number of friction sources on the particle surface.
	We are using the parallel $D_\parallel / D_0^S$, perpendicular $D_\perp / D_0^S$ and rotational $D_r / D_r^S$ diffusion coefficients obtained by the above method as input for our \ac{BD} simulations.
	The free single-sphere $(L^\ast = 0)$ translational $D_0^S$ and rotational $D_r^S$ diffusivities may be obtained from the respective Stokes-Einstein relations.
	The translational \ac{COM} diffusion coefficient of a free dumbbell is given by $D_0 = \frac{1}{3} D_\parallel + \frac{2}{3} D_\perp$.
	The Brownian time is accordingly given by $\tau = \sigma^2 / D_0^S$~\cite{Loewen1994, Kirchhoff1996}.

	In our simulations, the systems are subject to \acl{PBC} in all three Cartesian dimensions.
	The presented results have been obtained from \ac{BD} simulation runs with $N = 864$ particles in a constant volume $V$, so that the number density $\rho=N/V$. 
	The total simulation time is $100 \tau$ with the time step of $\Delta t = 10^{-4}\tau$.
	The \ac{fcc} ordered \acp{COM} with directors pointing in the $(1,1,1)$ directions have been set as initial configuration.
	To ensure equilibrium conditions results have only been obtained for $t \geq 50 \tau$. 
	All statistical correlations have been calculated for $50 \tau < t < 100\tau$ using every $100$-th time step.

	\begin{figure}[ht]
		\centering
		\includegraphics[]{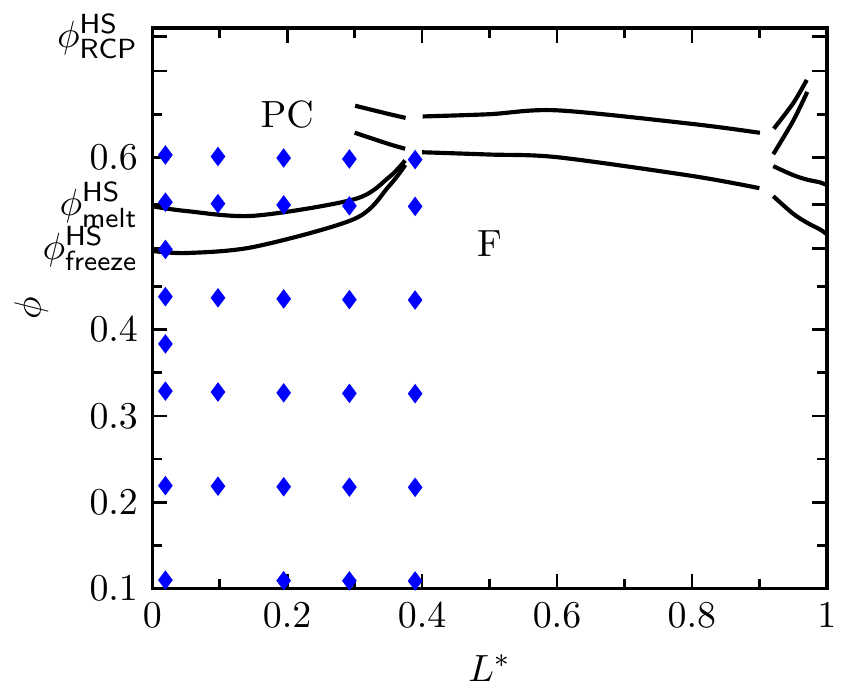}
		\caption{Schematic representation of the phase diagram of hard dumbbells~\cite{Marechal2008}. We investigate systems in the fluid (\ac{F}) and the \acf{PC} phase. The filled diamonds ($\color{blue}{\blacklozenge}$) indicate the state points investigated in this work, according to a Barker-Henderson mapping~(\ref{eq:diameter}) of the site-site Yukawa pair potential~(\ref{eq:potential}) onto hard spheres.}
		\label{fig:Marechal2008_phasediagram}
	\end{figure}

\subsection{Static structure}
	The expansion of the full radial distribution function~\cite{Schneider1986} yields two partial correlation functions of particular interest:
	We calculate the radial distribution function of the \acfp{COM}
	\begin{align}
		g(r) &= \frac{1}{\rho N}\left<\sum\limits^N_{i \neq j} \delta\left(\vec{r} - \vec{R}_{ij} \right)\right>,
	\end{align}
	and the orientational radial distribution function
	\begin{align}
		g_{P_2}(r) &= \frac{1}{\rho N g(r)} \left<\sum\limits^N_{i \neq j} P_2\left(\cos\theta_{ij}\right)\delta\left(\vec{r} - \vec{R}_{ij}\right)\right>,
	\end{align}
	where $\vec{R}_{ij}$ denotes the \ac{COM} separation vector of the $i$-th and $j$-th particles, and $\theta_{ij}$ is the angle between the respective directors. 
	The orientational distribution function $g_{P_2}(r)$ describes the spatial correlation of the particle directors with $P_2$ denoting the second Legendre polynomial.
	The function $g_{P_2}(r)$ equals $1$ for parallel aligned particles, $-\frac{1}{2}$ for perpendicular alignment, and vanishes for uncorrelated orientations.
	
\subsection{Dynamical properties}
	\subsubsection{Translational diffusion}
	We calculate the translational diffusion in the linear response regime from Green-Kubo-type of relations.
	In the limit of linear response, we thus obtain the long-time \ac{COM} self-diffusion coefficient ${D}_s^L$ from 
	the time correlation functions in equilibrium from the integral of the \ac{VACF} as 
	\begin{align}
		{D}(t) &= \int\limits^t_0 \left<\vec{V}(0)\cdot\vec{V}(t^\prime)\right> \upd t^\prime,
		\label{eq:diffoftime_vacf}
	\end{align}
	where $\vec{V}(t)$ is the instantaneous \ac{COM} velocity.
	The long-time limit of this function is the long-time self-diffusion coefficient
	\begin{align}
		{D}_s^L &= \lim_{t\rightarrow\infty} {D}(t).
		\label{eq:dinfvslstar_vacf}
	\end{align}

	\subsubsection{Orientational relaxation}
	The linear response orientational relaxation is quantified in terms of the \ac{DACF} defined as
	\begin{align}
		C_{\vec{\Omega} }(t) &= \left<\vec{\Omega}(0) \cdot \vec{\Omega}(t)\right> = e^{-2 D_r(t) t}.
	\end{align}
	We assume that the particle orientations become diffusive for long times with the constant $D_r^L$, therefore asymptotically the following relation holds~\cite{Kirchhoff1996}:
	\begin{align}
		C_{\vec{\Omega}}(t) &= e^{-2 D_r^L t}.
	\end{align}
	
	\subsubsection{Shear viscosity}
	In the limit of zero strain, we may obtain the dynamic viscosity of the suspension from equilibrium fluctuations.
	In particular, we calculate the relative dynamic viscosity difference from the off-diagonal \ac{SACF}
	\begin{align}
		z_{\alpha\beta}(t) &= \frac{1}{V \kB T} \left<\sigma_{\alpha\beta}(0) \sigma_{\alpha\beta}(t)\right>,
		\label{eq:sacf}
	\end{align}
	where the symmetric stress tensor is defined as
	\begin{align}
		\sigma_{\alpha\beta} &= \frac{1}{V}\sum_{i = 1}^{N} \sum_{j>i}^{N} {r}_{ij\alpha}F_{ij\beta},
		\label{eq:stress}
	\end{align}
	and $\alpha \neq \beta$ denote Cartesian components. 
	From our \ac{BD} simulations, the potential part of the stress tensor is available, therefore (\ref{eq:stress}) does not contain a momentum part.
	
	The steady shear viscosity difference is obtained from the integral of the \ac{SACF} as
	\begin{align}
		\eta_0 &= \int\limits_0^\infty z_{\alpha\beta}(t) \upd t.
	\end{align}
	
\section{Results and discussion}
\subsection{Spatial structure}
\begin{figure}[ht]
		\centering
		\includegraphics[]{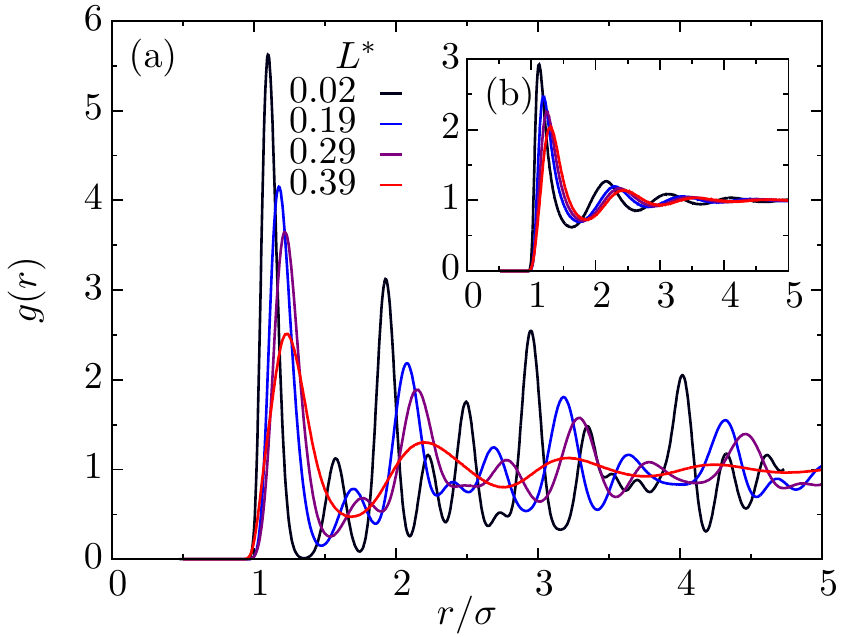}
			\caption{The radial distribution function $g(r)$ for the dumbbell center-of-mass at a volume fraction $\phi = 0.60$: black, blue, and purple are in the \ac{PC}, and red in the \ac{F} phases, respectively. Inset: The same for a packing fraction $\phi = 0.44$, where all systems are in the \ac{F} phase.}
		\label{fig:rdf_g_com_phi0.55_0.40}
	\end{figure}

	In Fig.~\ref{fig:rdf_g_com_phi0.55_0.40} the \ac{COM} \acfp{RDF} of dumbbell suspensions at the volume fraction $\phi = 0.60$ are displayed for various elongations.
	The signature of the phase transition from \ac{PC} to \acf{F} may be readily observed here.
	At state points with $L^\ast \lesssim 0.3$ the system is in the \ac{PC} phase, cf. Fig.~\ref{fig:Marechal2008_phasediagram}, where
	the \ac{COM} positions are frozen in a crystal lattice and the directors do not show any long-distance correlations.
	Hence, the corresponding \acp{RDF} show strong correlations and long-ranged oscillations. They vanish at $L^\ast = 0.39$, where the system is in the isotropic fluid phase.
	The inset of Fig.~\ref{fig:rdf_g_com_phi0.55_0.40} shows the \acf{COM} \acfp{RDF} $g(r)$ at the fixed colloidal packing fraction $\phi = 0.44$. 
	All shown curves are in the \acf{F} phase. 	With increasing elongation, the extrema get less pronounced and are shifted to greater distances due to the larger effective particle sizes.

\begin{figure}[ht]
		\centering
		\includegraphics[]{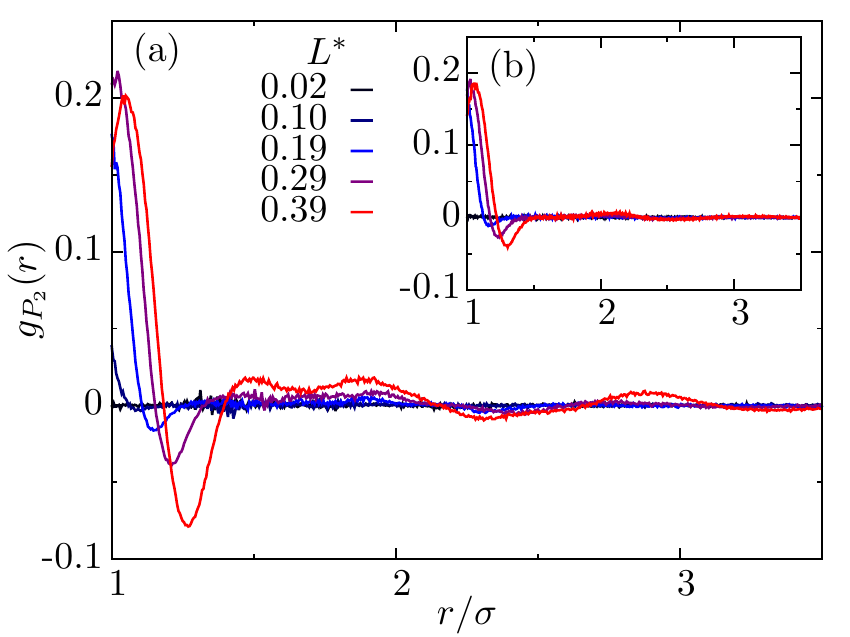}
		\caption{Orientational pair correlation function $g_{P_2}(r)$ at a volume fraction $\phi = 0.60$: the black, blue, and purple curves present data for the systems in the \acf{PC} phase and red in the \acf{F} phase, respectively. Inset: The same for a packing fraction $\phi = 0.44$ where all systems are in the \acl{F} phase.}
		\label{fig:rdf_gp2_phi0.55_0.40}
	\end{figure}

	The orientational \acp{PCF}	$g_{P_2}$ at the volume fractions $\phi = 0.60$ and $\phi = 0.44$ are depicted in Fig.~\ref{fig:rdf_gp2_phi0.55_0.40} and its inset, respectively, for various representative elongations.
	As expected, the orientational correlation in space is essentially flat for suspensions of nearly spherical particles ($L^\ast = 0.02$) at all volume fractions.
	At roughly $r/\sigma \simeq 1$, the $g_{P_2}$ for non-zero elongations show maxima indicating a strong correlation of nearest neighbors at contact. 
	This first correlation peak moves slightly away from $r = \sigma$ as the elongation increases.
	At the first maxima, the functions are positive, representing a preferably parallel orientation of dumbbell particles at close contact.
	For a bit larger distances, $r/\sigma \simeq 1.1-1.3$, negative dips indicating orientational anti-correlations are observed which tend to higher distances as the elongation is increased.
	Hence, in the \ac{PC} phase at $\phi = 0.60$, there are non-vanishing next neighbor correlations which grow with elongation.
	With further increasing aspect ratio, the orientational correlation becomes non-zero over a higher distance as the system crosses to the dense fluid phase at $\phi = 0.44$. Here, we observe a new correlation peak at in the dense fluid state for $L^\ast = 0.39$ close to $r/\sigma \simeq 1.8$ and 2.8, indicating growing second and third neighbor correlations. Thus, in the fluid phase the correlations are longer ranged due to the higher disorder and collisions when compared to the \ac{PC} phase.
				
\subsection{Translational diffusion}

	\subsubsection{Time-dependent diffusion coefficient}
		
	\begin{figure}[ht]
		\centering
		\includegraphics[]{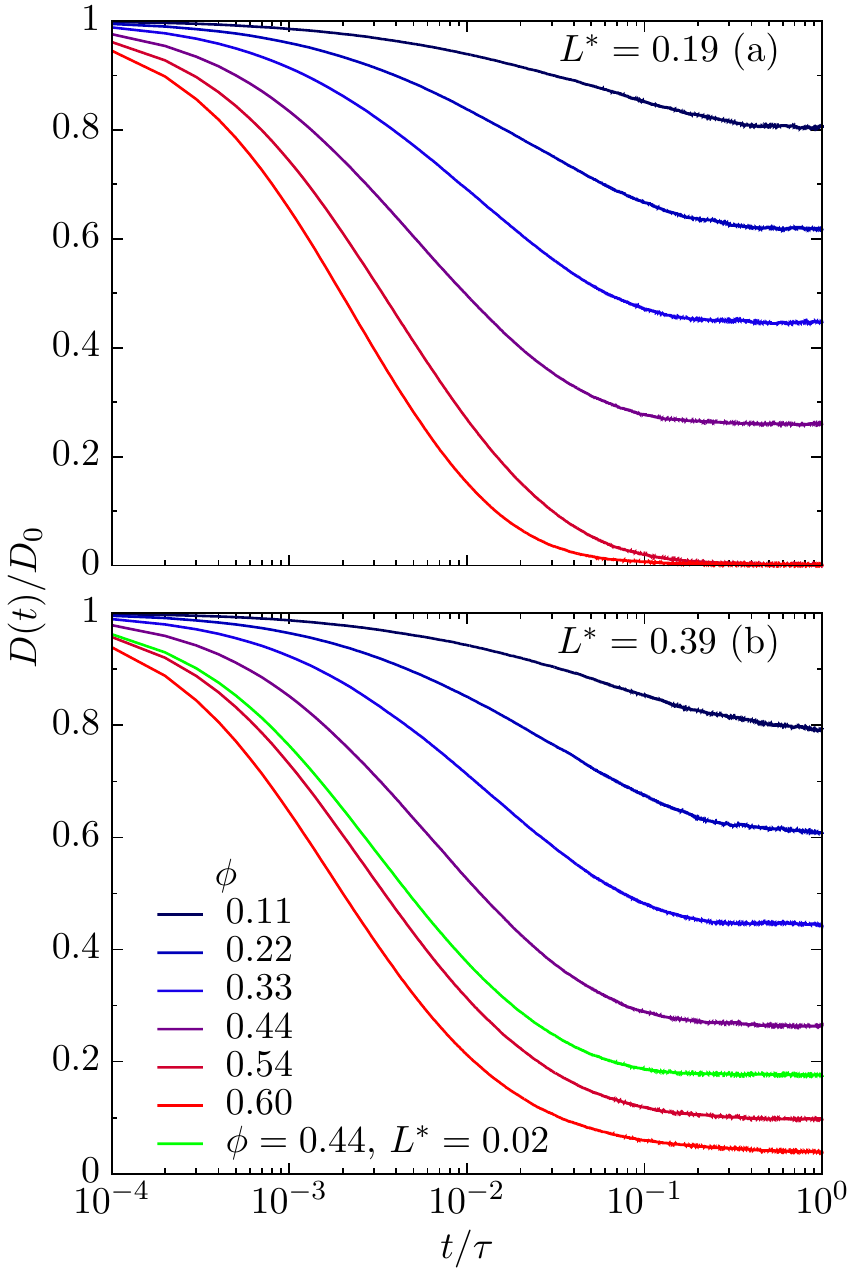}
		\caption{Time-dependent diffusion coefficient ${D}(t)/D_{0}$ from the \ac{VACF} at (a) $L^\ast = 0.19$ and (b) $L^\ast = 0.39$ for different packing fractions. The dashed line in (b) is at state point $\phi=0.44, L^\ast=0.02$ for a direct comparison.}
		\label{fig:diffoftime_vacf_lstar0.20_lstar0.40}
	\end{figure}

	Figure~\ref{fig:diffoftime_vacf_lstar0.20_lstar0.40}~a) and b) show the time-dependent \ac{COM} diffusion coefficients $D(t)$ at constant elongation $L^\ast = 0.19$ and $L^\ast = 0.39$ for different colloidal volume fractions, respectively. 
	At early times ($t\lesssim 10^{-4}$), before the particles feel the interacting neighbors in a non-dilute suspension, 
	the diffusion coefficients are close to the short-time limits of a single, free particle.	
	Within times $t\lesssim 10^{-2}\tau$, a continuous transition to long-time diffusion starts which is reached at about the Brownian time scale $\tau$.
	With increasing volume fraction $\phi$, the cross-over from short-time to long-time behavior sets in earlier and is steeper.
	At the elongation $L^\ast = 0.19$ the systems with packing fractions $\phi = 0.54$ and $\phi= 0.60$ are in the \ac{PC} phase. Due to the translational constraints in the crystal, the diffusion vanishes at about a tenth of a Brownian time.
	The comparison of the diffusion between $L^\ast = 0.19$ and $L^\ast = 0.39$ at $\phi = 0.44$ in the dense fluid in panel (a) 
	demonstrates that the influence of the aspect ratio is very small in the range of focus of this study. 			
	Since all the shown curves for the elongation $L^\ast = 0.39$ are in the \ac{F} phase (panel (b)), 
	the diffusion coefficients are not vanishing for all volume fractions. 
	
	\begin{figure}[ht]
		\centering
		\includegraphics[]{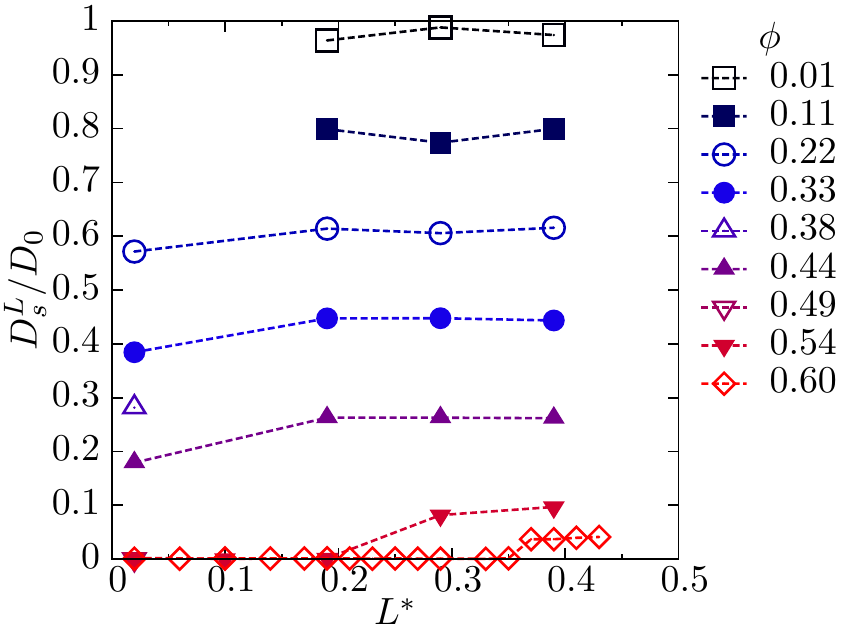}
		\caption{Long-time self-diffusion coefficients ${D}_s^L$ obtained from the \acp{VACF} via relation (\ref{eq:diffoftime_vacf}).}
		\label{fig:dinfvslstar_vacf}
	\end{figure}
		
	In Fig.~\ref{fig:dinfvslstar_vacf} the values of the long-time \ac{COM} self-diffusion coefficients $D_s^L$ are presented for various elongations and packing fractions.
	As expected, the diffusion is significantly decreased for larger packer fractions, as known for the spherical reference case.
	Surprisingly, the dependence of the diffusion on elongation is weak and we find that this is the case for both parallel and perpendicular long-time diffusion as well. 
	Apparently, the anisotropy is not large enough to substantially change the long-time (single colloid) friction in these systems.
	Furthermore, the data shows a discontinuous transition to vanishing diffusion at elongations where the transition to the \ac{PC} phase takes place, see the curves for the large packing fractions $\phi \geq 0.54$.

\subsection{Rotational relaxation and diffusion}

\begin{figure}[ht]
	\centering
	\includegraphics[]{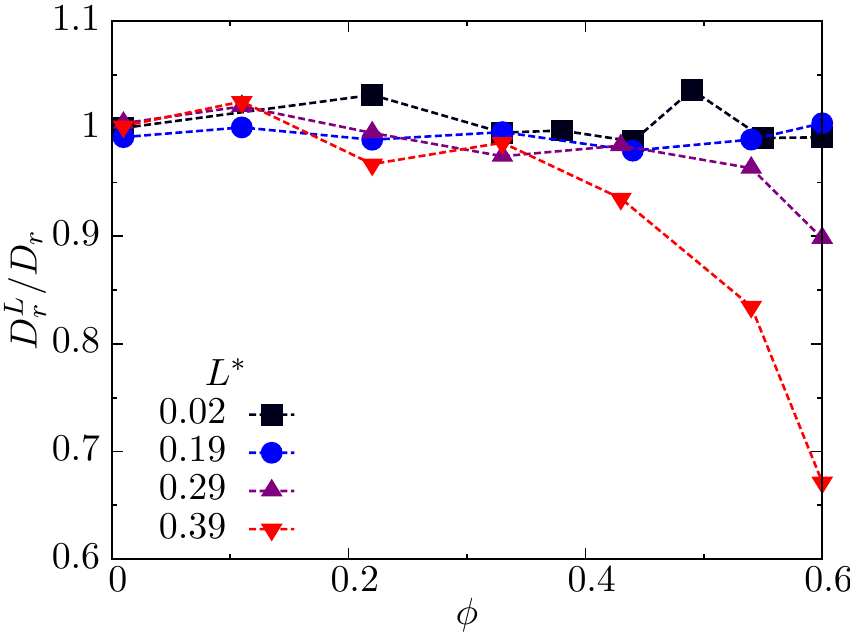}
	\caption{Dependence of rotational diffusion coefficients $D_r^L$ at various elongations from exponential fits to \ac{DACF} on volume fraction $\phi$. The statistical error of the data is estimated to be $\pm 5 \,\%$, calculated from block averages over independent trajectories of length of $10\tau$.}
	\label{fig:drotinfvsphi_dacf}
\end{figure}

Fig.~\ref{fig:drotinfvsphi_dacf} shows the rotational diffusion coefficients obtained from exponential fits to the \acp{DACF}. 
At small elongations $L^\ast \lesssim 0.2$ the rotational diffusion is almost constant with respect to volume fraction, albeit the system crosses from the \acl{F} to the \ac{PC} phase at high fractions.
Hence, for these small elongations the particles indeed rotate almost freely in the \ac{PC} phase.
The obvious statistical outliers in the rotational diffusion data for small packing fractions $\phi \lesssim 0.3$ are within the statistical error of $\pm 5\%$, estimated from block averages over independent trajectories of length of $10\tau$.
However, suspensions of dumbbells with $L^\ast = 0.29$ show a notable drop of the rotational diffusion coefficients for densities higher than $\phi = 0.5$ and packing effects clearly influence the rotation correlations in the \ac{PC} phase. 
 On increasing the elongation further, now a clear non-linear density-dependence of the rotational relaxation emerges which becomes obvious for $L^\ast = 0.39$ for packings larger than $\phi \gtrsim 0.3$. Here, the system is in the fluid phase only and the closer contacts in the disordered systems alter the rotational dynamics substantially.
 At the highest investigated packing fraction ($\phi = 0.60$) the rotational diffusion drops down by more than $30 \,\%$. 
		
\subsection{Shear viscosity}

\begin{figure}[ht]
	\centering
	\includegraphics[]{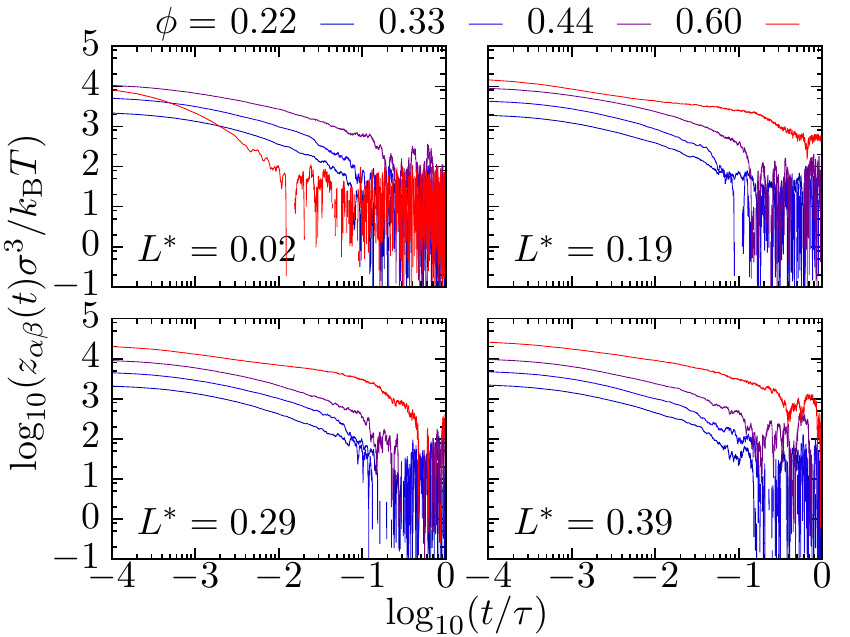}
	\caption{\acp{SACF} at different elongations (increasing from top left to bottom right panels) and volume fractions.}
	\label{fig:sacf_mult_lstar}
\end{figure}

Figure~\ref{fig:sacf_mult_lstar} shows the stress autocorrelation function \acp{SACF} at various aspect ratios $L^\ast$. 
For the smaller packing fractions $\phi < 0.5$, the correlation functions decay slower for higher densities while
the elongation dependence is rather weak. However, at the volume fraction $\phi = 0.60$ the system is in 
the \ac{PC} phase for the shown elongations, except for $L^\ast = 0.39$. Here an interesting behavior can be observed: 
at $L^\ast = 0.02$ the \ac{SACF} decays more rapidly in the \ac{PC} state point than in the fluid states.
However, for the next two larger elongations $L^\ast = 0.19$ and $L^\ast = 0.29$, the \ac{SACF} decays slower and develops a new slow time scale for times about 10$^{-1}\tau$ in the \ac{PC} phase.
This must be assigned to slow stress relaxations in the dense crystal phase likely stemming from more pronounced translational-rotational couplings~\cite{Ruth}, since they are absent for the small elongations at the same density.
This long-time tail occurring in the \ac{PC} phase seems to diminish when 
crossing into the \ac{F} phase by increasing the aspect ratio to $L^\ast = 0.39$.
	
\begin{figure}[ht]
	\centering
	\includegraphics[]{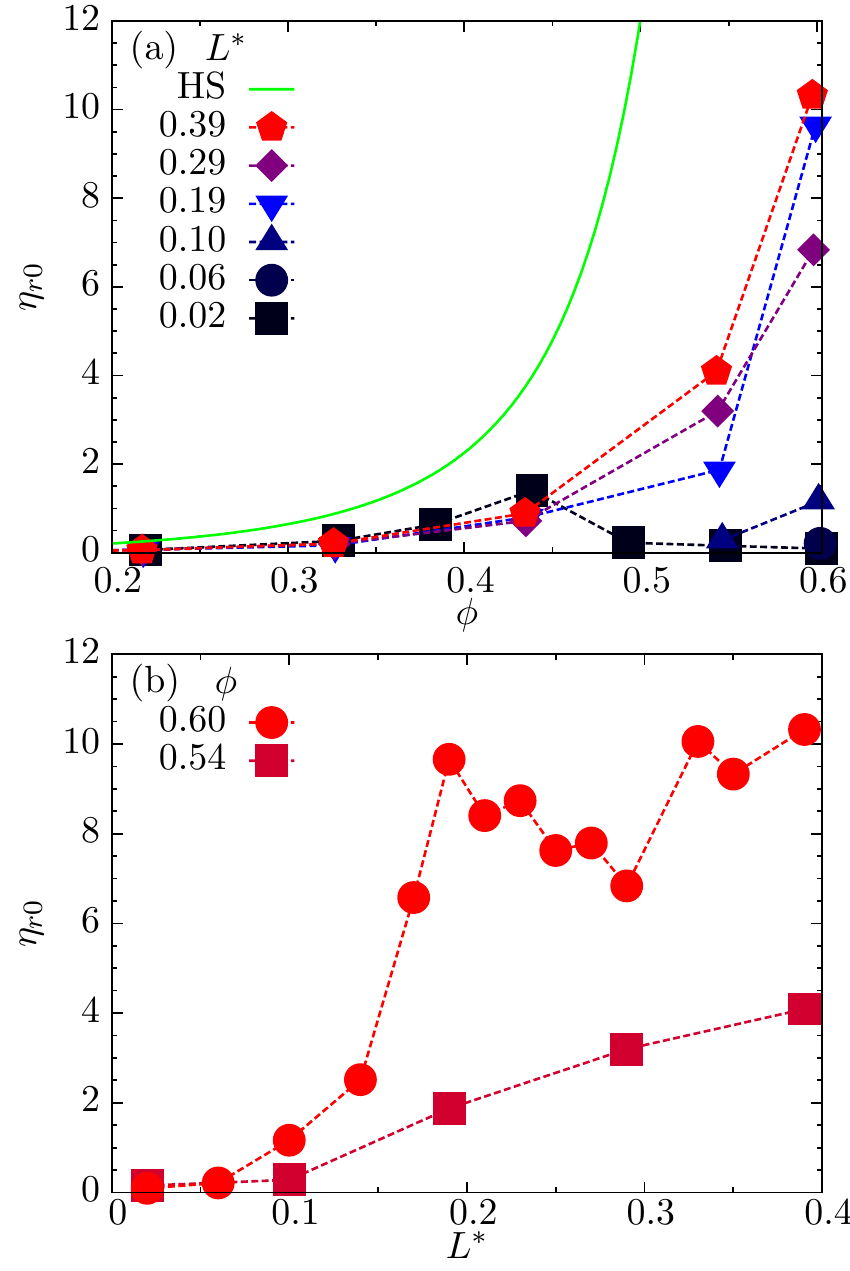}
		\caption{Steady shear relative viscosity difference $\eta_{r0}$ (a) as a function of the volume fraction $\phi$ and (b) versus elongation $L^\ast$ at high densities. The \acf{HS} line ($\color{green}{-}$) refers to the phenomenological Krieger-Dougherty theory~\cite{Heyes1994, Russel1992}.}
		\label{fig:etavsphi}
	\end{figure}

Figure~\ref{fig:etavsphi} a) shows the packing fraction dependence of the relative steady shear viscosity $\eta_{r0} = \eta_0 / \eta_s$ for different elongations $L^\ast$, with $\eta_s$ being the viscosity of the suspending medium.
These data are obtained from integration of the \acp{SACF} in Fig.~\ref{fig:sacf_mult_lstar}.
As an important comparison, also the \acf{HS} reference data following the empirical Krieger-Dougherty relation~\cite{Heyes1994, Russel1992} for the \emph{purely fluid} phase is shown.
As we readily see, there is a difference of the viscosity for a fluid of \aclp{HS} when compared to our minimal elongation data (at $L^\ast = 0.02$).
This can be understood as a result of the sensitivity of viscosities to the softness of the chosen pair potential for repulsive spheres~\cite{Heyes2005a}.
In the present case, the $L^* = 0.02$ system with minute anisotropy can be considered equal to the purely spherical system.
At higher packing the \ac{HS} system is in the crystal phase and the linear shear response drops to zero (and not accounted for in the fluid-state Krieger-Dougherty approach). 
The shear viscosity of dumbbells is similar to the \ac{HS} system in the fluid phase below packings of $\phi\lesssim 0.45$ but deviates
strongly for larger packings for all elongations where the dumbbell shear response substantially increases with $L^*$. 
We explain the increase by the long time tails in the \acp{SACF} in Figure~\ref{fig:sacf_mult_lstar} due to the relaxation of the rotational degrees of freedom missing in the \ac{HS} case.
The large increase of the shear response is somewhat unexpected since the translational degrees of freedom are still frozen, except for the largest elongation $L^\ast = 0.39$ for which all shown data are in the fluid phase.
We note that the error bars of this data are hard to estimate due to the long-time tails in the \acp{SACF}.
While we believe from the systematic behavior of the data that the trends are reliable, small fluctuations of the data as for $\phi = 0.44$ are very probable within the statistical uncertainty.

	Figure~\ref{fig:etavsphi} b) shows the dependence of the relative steady shear viscosity on the elongation at the high packing fractions $\phi\geq 0.39$.
	This view reveals a transition from the hard spherical case of almost vanishing viscosity to a finite level at about a critical elongation of about $L^*=0.15$. 
	This is most evident at the highest packing fraction where the transition is relatively steep.
	(The fluctuations of the data between $L^\ast = 0.2$ and 0.3 are most certainly within the statistical uncertainty.)
	Hence, it may be inferred from the equilibrium linear response data that a small deviation from the spherical shape already to an aspect ratio of $L^*\gtrsim 0.15$ induces a dramatic collective response that manifests itself in a large increase of the linear response shear viscosity. 
		
\section{Summary and concluding remarks}
In summary, we have investigated the equilibrium structure and fluctuations of colloidal dumbbells by means of Brownian Dynamics computer simulations.
We focused on high packing fractions and on the weak elongation regime ($L^*<0.4$) 
where the dumbbells are predominantly in the plastic crystal phase. Our systematic investigation revealed the expected structural 
changes with larger elongation with respect to the hard sphere reference case and very localized orientational correlations, 
typically just involving next-neighbor correlations. These relatively weak correlations are also reflected in only minor changes
in the translational and rotational diffusion coefficients for most of the investigated elongations, except for the highest 
elongation in the fluid phase where the rotational diffusion drops by ca. $30 \,\%$ when compared to the free rotation. 
However, the linear response shear viscosity exhibits a dramatic increase even in the plastic crystal phase at high packing fractions ($\phi\gtrsim 0.5$) beyond a critical elongation of about $L^*=0.15$. This result is surprising in view of the relatively weak effects of elongation found before. 
Here one should rationalize that the linear response shear viscosity expresses collective relaxation time scales and not (more local) single molecule dynamic properties.
Apparently beyond a critical, surprisingly small anisotropy, newly occurring rotational-translational couplings must be made responsible for the slow time scales appearing at higher elongations in the crystal~\cite{Ruth}.
Hence, more detailed calculations and modeling of these rotational-translational correlations in weakly anisotropic model 
systems shall be interesting for future studies, in particular beyond linear response where probably more substantial 
dynamical effects of increased anisotropy may occur~\cite{Chu2014}.

\section{Acknowledgments}
The authors thank Matthias Ballauff and Fangfang Chu for helpful discussions. JD would like to express his gratitude to Jean-Pierre Hansen for being a great teacher and mentor, the plenty of inspiring scientific exchanges, and his kind hospitality in Cambridge and Paris.


\begin{thebibliography}{38}
\providecommand{\url}[1]{\texttt{#1}}
\providecommand{\urlprefix}{URL }
\markboth{Taylor \& Francis and I.T. Consultant}{Molecular Physics}

\bibitem{HansenMcDonald}
J. Hansen and I. McDonald, \emph{Theory of Simple Liquids}, 3rd ed.   (Academic
  Press, London, 2006).

\bibitem{Pusey:Nature}
P.N. Pusey and W. van Megen,  Nature  \textbf{320}, 340 (1986).

\bibitem{LesHouches}
P.N. Pusey, Colloidal Suspensions. in \emph{Les Houches Summer School: Liquid,
  Freezing and Glass Transition}, edited by J.-P. Hansen, D. Levesque and J.
  Zinnjustin.

\bibitem{Loewen2001}
H. L{\"o}wen,  J. Phys.: Condens. Matter.  \textbf{13}, R415 (2001).

\bibitem{Ackerson1990a}
B.J. Ackerson,  J. Rheol.  \textbf{34}, 553 (1990).

\bibitem{Haw1998}
M.D. Haw, W.C.K. Poon and P.N. Pusey,  Phys. Rev. E  \textbf{57}, 6859 (1998).

\bibitem{Johnson}
P.M. Johnson, C.M. van Kats and A. van Blaaderen,  Langmuir  \textbf{21}, 11510
  (2005).

\bibitem{Mock2007}
E.B. Mock and C.F. Zukoski,  J. Rheol.  \textbf{51}, 541 (2007).

\bibitem{Mock2007a}
E.B. Mock and C.F. Zukoski,  Langmuir  \textbf{23}, 8760 (2007).

\bibitem{Kramb2010}
R.C. Kramb, R. Zhang, K.S. Schweizer and C.F. Zukoski,  Phys. Rev. Lett.
  \textbf{105}, 055702 (2010).

\bibitem{Kramb2011}
R.C. Kramb and C.F. Zukoski,  J. Rheol.  \textbf{55}, 1085 (2011).

\bibitem{Kramb2011a}
R.C. Kramb and C.F. Zukoski,  J. Rheol.  \textbf{55}, 1069 (2011).

\bibitem{Forster2011}
J.D. Forster, J.G. Park, M. Mittal, H. Noh, C.F. Schreck, C.S. O'Hern, H. Cao,
  E.M. Furst and E.R. Dufresne,  ACS Nano  \textbf{5}, 6695 (2011).

\bibitem{wagner}
M.M. Panczyk, J.G. Park, N.J. Wagner and E.M. Furst,  Langmuir  \textbf{29}, 75
  (2013).

\bibitem{Hoffmann2008}
M. Hoffmann, Y. Lu, M. Schrinner, M. Ballauff and L. Harnau,  J. Phys. Chem. B
  \textbf{112}, 14843 (2008).

\bibitem{Nagao2010}
D. Nagao, C.M. van Kats, K. Hayasaka, M. Sugimoto, M. Konno, A. Imhof and A.
  van Blaaderen,  Langmuir  \textbf{26}, 5208 (2010).

\bibitem{Peng2012}
B. Peng, H.R. Vutukuri, A. van Blaaderen and A. Imhof,  J. Mater. Chem.
  \textbf{22}, 21893 (2012).

\bibitem{Chu2012}
F. Chu, M. Siebenb\"{u}rger, F. Polzer, C. Stolze, J. Kaiser, M. Hoffmann, N.
  Heptner, J. Dzubiella, M. Drechsler, Y. Lu and M. Ballauff,  Macromol. Rapid
  Commun.  \textbf{33}, 1042 (2012).

\bibitem{Ruth}
R.~M. Lynden-Bell and K. H. Michel, Rev. Mod. Phys. \textbf{66}, 721 (1994).

\bibitem{Vega1992}
C. Vega, E.P.A. Paras and P.A. Monson,  J. Chem. Phys.  \textbf{97}, 8543
  (1992).

\bibitem{Vega1992a}
C. Vega, E.P.A. Paras and P.A. Monson,  J. Chem. Phys.  \textbf{96}, 9060
  (1992).

\bibitem{Vega1997}
C. Vega and P.A. Monson,  J. Chem. Phys.  \textbf{107}, 2696 (1997).

\bibitem{Marechal2008}
M. Marechal and M. Dijkstra,  Phys. Rev. E  \textbf{77}, 061405 (2008).

\bibitem{Ni2011}
R. Ni and M. Dijkstra,  J. Chem. Phys.  \textbf{134}, 034501 (2011).

\bibitem{Paras1992}
E.P.A. Paras, C. Vega and P.A. Monson,  Mol. Phys.  \textbf{77}, 803 (1992).

\bibitem{Chu2014}
F. Chu, N. Heptner, Y. Lu, M. Siebenb\"{u}rger, P. Lindner, J. Dzubiella and M.
  Ballauff,  Langmuir   (2015), Article ASAP, DOI:10.1021/la504932p.

\bibitem{Barker1967}
J.A. Barker and D. Henderson,  J. Chem. Phys.  \textbf{47}, 4714 (1967).

\bibitem{Heyes1994}
D.M. Heyes and P.J. Mitchell,  J. Phys.: Condens. Matter.  \textbf{6}, 6423
  (1994).

\bibitem{Phung1996}
T.N. Phung, J.F. Brady and G. Bossis,  J. Fluid Mech.  \textbf{313}, 181
  (1996).

\bibitem{Carrasco1999}
B. Carrasco and J.G. de~la Torre,  Biophys. J.  \textbf{76}, 3044 (1999).

\bibitem{delaTorre2002}
J.G. de~la Torre and B. Carrasco,  Biopolymers  \textbf{63}, 163 (2002).

\bibitem{delaTorre2007}
J.G. de~la Torre, G.D. Echenique and A. Ortega,  J. Phys. Chem. B
  \textbf{111}, 955 (2007).

\bibitem{Diaz1989}
F.G. Diaz, J.G. de~la Torre and J.J. Freire,  Polymer  \textbf{30}, 259 (1989).

\bibitem{Tirado1984}
M.M. Tirado, C.L. Martinez and J.G. de~la Torre,  J. Chem. Phys.  \textbf{81},
  2047 (1984).

\bibitem{Loewen1994}
H. L{\"o}wen,  Phys. Rev. E  \textbf{50}, 1232 (1994).

\bibitem{Kirchhoff1996}
T. Kirchhoff, H. L{\"o}wen and R. Klein,  Phys. Rev. E  \textbf{53}, 5011
  (1996).

\bibitem{Schneider1986}
J. Schneider, W. Hess and R. Klein,  Macromolecules  \textbf{19}, 1729 (1986).

\bibitem{Russel1992}
W.B. Russel, D.A. Saville and W.R. Schowalter, \emph{Colloidal Dispersions}
  Cambridge Monographs on Mechanics and Applied Mathematics  (Cambridge
  University Press (CUP), Cambridge, 1992).

\bibitem{Heyes2005a}
D.M. Heyes and A.C. Bra\'{n}ka,  J. Chem. Phys.  \textbf{122}, 234504 (2005).

\end{thebibliography}
\end{document}